\documentclass[epj]{svjour}  

\usepackage{amsmath,graphicx}
\usepackage{amsmath} 
\usepackage{amssymb, bbm}
\begin{document} 
\title{Effects of order $\alpha^3$ on the determination of the Pauli form factor $F_2$ of the $\tau$-lepton}

\author
{F. Krinner\inst{1} \and N. Kaiser\inst{2}}
\institute{$^1$Max Planck Institut f\"ur Physik, 80805 
M\"unchen\\ $^2$Physik-Department T39, Technische 
Universit\"{a}t M\"{u}nchen}

\abstract{
 We introduce optimal observables to measure the Pauli form factor $F_2$ of the $\tau$-lepton in the pair-production process ${\rm e}^- {\rm e}^+ \to \tau^-\tau^+$ from the intensity distribution of the decay products. The spin-density matrix for the production process is calculated in QED up to order $\alpha^3$ including virtual photon-loops and soft bremsstrahlung, as well the $\gamma Z^0$ interference. We find that the decay channel $(\rho^-\nu_\tau)\times (\rho^+ \bar\nu_\tau)$ yields the best resolution for Re$F_2(s)$ and Im$F_2(s)$ due to its high branching fraction. We also study the bias that is introduced in the determination  of $F_2$, if the production spin-density matrix is taken in tree-level (one-photon exchange) approximation.
  \PACS{
     {}{}  } }
\maketitle

\section{Introduction}\label{intro}
In the light of the very accurate results for the anomalous magnetic moment  
$a_\mu = (g_\mu-2)/2$ of the muon presented recently in Ref.~\cite{muon}, a measurement of the analogous  quantity for the $\tau$-lepton would clearly be of great interest. 
In quantum electrodynamics (QED) the coupling vertex of  the photon to a $\tau$-lepton
pair is  of the form $e\, \bar u_{\tau^-} \Gamma^\mu v_{\tau^+}$,
with $\bar u_{\tau^-} $ and $v_{\tau^+} $ free Dirac-spinors and the composed Dirac-matrix:
\begin{equation}\label{eq:QEDgamma}
\Gamma^\mu = F_1(s) \gamma^\mu + \frac{i}{2m_\tau}{F_2(s)}\sigma^{\mu\nu} q_\nu
\,, \quad \sigma^{\mu\nu}={i \over 2}[\gamma^\mu ,\gamma^\nu]\,, \end{equation}
where $q^\mu$ is the four-momentum carried by the virtual photon and $s = q^\mu q_\mu$.
Here, one has introduced the so-called Dirac form factor $F_1(s)$, normalized to $F_1(0)=1$, and the so-called Pauli form factor $F_2(s)$, that gives the anomalous magnetic moment  $F_2(0) = a_\tau = (g_\tau - 2)/2$.

One way to access this coupling is the production of 
$\tau$-lepton pairs at ${\rm e}^-{\rm e}^+$-colliders:
\begin{equation}
\rm{e}^- + e^+ \to \tau^- + \tau^+\,.\end{equation}
In such a process, a non-zero real part Re$F_2(s)$ leads 
to a modification of the cross section, while a non-zero  imaginary part
Im$F_2(s)$ even alters the spin-correlation between the produced $\tau^-$ and 
$\tau^+$ \cite{bernabeu}. However, due to the very short life-time (of less than a picosecond), 
$\tau$-leptons typically decay before they can reach any detector and therefore only their spin correlation can be accessed through the angular distribution of the decay products. At the same time, one should keep in mind, that in  $\tau$-pair production only the kinematical region $s>4m_\tau^2$ is 
accessible, and thus one does not measure the anomalous magnetic moment $a_\tau$ directly.

While a similar measurement aiming at  the electric dipole moment 
$d_\tau$ has been performed \cite{belle}, no such analysis was done for 
$a_\tau$ because of expected contaminations of the measurement by 
higher order QED effects. Since the $\gamma \tau^-\tau^+$ coupling that incorporates the electric dipole moment $d_\tau$ stems from a CP-odd higher dimensional operator, these QED corrections have been assumed to be of no relevance for the pertinent CP-odd observable.
It has been argued in the past that a 
measurement of $F_2(s)$ in $\tau$-pair production events may be 
meaningful and feasible, e.g. due to suppression of the contaminations at $\sqrt{s} = m_{\Upsilon(4S)}=10.58\,$GeV. For this reason we want to determine  in this work the actual size 
of these contaminations by comparing results from leading order QED 
with next-to-leading order QED calculations, and relate these corrections to
the resolution that can be achieved in an extraction of $F_2(s)$.

Our paper is organized as follows. In section~\ref{sec:OO} we 
introduce our analysis method employing the so-called optimal 
obervables. In section~\ref{sec:NLO} we derive analytical expressions 
for the  spin-density matrix of $\tau$-pair production including virtual photon-loops and soft bremsstrahlung up to  order $\alpha^3$, and we estimate the 
effect of these next-to-leading order corrections in section~\ref{sec:studies}. 
Finally, the conclusions of our study are given in section~\ref{sec:conclusions}. Appendix A collects the basic loop-functions that are encountered in the evaluation of the
two-photon exchange box diagrams, and in appendix B we present our results based on optimal observables concerning the measurement of the electric dipole moment $d_\tau$.  

\section{Pauli form factor $F_2(s)$ and optimal observables}
\label{sec:OO}
In order to access the Pauli form factor $F_2(s)$, we start from the full intensity distribution ${\cal I}$ for $\tau^-\tau^+$ pair production, including their weak decays. It involves four sums over helicities and takes the form:
\begin{equation}
{\cal I}= \sum_{\lambda_\mp,\lambda_\mp^{(\prime)}} \chi^{\rm prod}_{\lambda_-\lambda_+,\lambda_-^\prime\lambda_+^\prime}\cdot \chi_{\lambda_-,\lambda_-^\prime}^- \cdot \chi_{\lambda_+,\lambda_+^\prime}^+\,,\end{equation}
where $\chi_{\lambda_-,\lambda_-^\prime}^-$ and $\chi_{\lambda_+,\lambda_+^\prime}^+$ stand for the spin-density matrices of the respective weak $\tau^\mp$-decays, 
and $\chi^{\rm prod}_{\lambda_-\lambda_+,\lambda_-^\prime\lambda_+^\prime}$ denotes the 
spin-density matrix of $\tau^-\tau^+$ pair-production in ${\rm e}^-{\rm e}^+$ annihilation, which we will give in analytical form up to order $\alpha^3$ (with $\alpha =1/137.036$) in 
section~\ref{sec:NLO}. The spin-density matrices $\chi^\mp_{\lambda_\mp,\lambda_\mp^\prime}$ for the weak decays arise from tree-level amplitudes involving $W^\mp$ exchange in the
case of leptonic decays, and for hadronic $\tau^\mp$-decays they can be found in Refs.~\cite{dumm1,dumm2}.
Since the Pauli form factor $F_2$ is small, one can expand the intensity distribution in terms of $F_2$ up to linear order:
\begin{equation}\label{eq:Mexpand}
\mathcal{I} = \mathcal{I}_0 + {\rm Re} F_2\, \mathcal{I}_{\rm Re} + {\rm Im}F_2\, \mathcal{I}_{\rm Im} \,.\end{equation}
Using this expansion, we define the following optimal 
observables \cite{OOref}:
\begin{equation}\label{eq:OOdef}
\mathcal{O}^{\rm opt}_{\rm Re,Im} = \frac{\mathcal{I}_{\rm Re,Im}}{\mathcal{I}_0}
,\end{equation}
which are exclusively sensitive to contributions from $F_2$. For a given set $\mathcal{S}$ of data, the value 
of $F_2$ is determined through the expectation values $\langle \mathcal{O}^{\rm opt}_{\rm Re,Im}\rangle_\mathcal{S}$ of the optimal observables. We generate 
Monte Carlo data sets in order to 
predict the dependence of the expectation values of the optimal observables on $F_2$. In contrast to previous analyses \cite{belle,chen}, we do not generate
data sets at various working points for $F_2$ and then perform a linear fit through these working points. Instead, we generate a single data set $\mathcal{S}_0$, using the value $F_2 = 0$ and determine the 
dependence of the expectation values using a reweighting approach.
For this purpose, we construct a weight $w(F_2)$ for every event that reweights  
the data set, such that it would correspond to a non-zero value $F_2\neq0$:
\begin{equation}\label{eq:weights}
w(F_2) = \big(\mathcal{I}_0 + {\rm Re}F_2\, \mathcal{I}_{\rm Re} + {\rm Im}F_2 \,\mathcal{I}_{\rm Im} + |F_2|^2\, \mathcal{I}_{|F_2|^2}\big)\big/\mathcal{I}_0
,\end{equation}
where quadratic terms  have to be kept, since these weights will enter non-linearly. Having these weights, we can predict the expectation value for an optimal observable with $F_2\neq 0$ using the formula:
\begin{equation}\label{eq:reweight}
\langle \mathcal{O}^{\rm opt}_{\rm Re,Im}\rangle_{F_2\neq 0} = \langle w(F_2)\mathcal{O}^{\rm opt}_{\rm Re,Im}\rangle_{F_2=0} /\langle w(F_2)\rangle_{F_2 = 0}
\,,\end{equation}
which results in two coupled quadratic equations that can be solved for Re$F_2$ and Im$F_2$. Note that since $F_2$ is small, one takes the solution with the smallest magnitude $|F_2|$ as the correct one.

\subsection{Reconstruction effects}
\label{sec:OObar}

The method described so far relies on the calculation of the intensity $\mathcal{I}$
from the kinematics of the decay products in an event. However, in reality this kinematics is not fully accessible since at least one neutrino escapes for every $\tau$-decay.
This results in a two-fold kinematical ambiguity if one single neutrino is produced in each weak decay, corresponding to semi-leptonic 
channels. Moreover, in leptonic decays there remains an ambiguity in three unknown kinematical variables, the invariant mass $m_{\nu_\tau\bar\nu_\ell}$ and the two angles describing the $\nu_\tau\bar\nu_\ell$  system.

In order to  account for this problem, we use the integrated intensity $\overline{\mathcal{I}}$, where all unknown kinematical 
degrees of freedom are integrated over and thus all ambiguities are averaged out. Using an expansion of 
$\overline{\mathcal{I}}$ analogous to Eq.(\ref{eq:Mexpand}), we define integrated optimal observables 
$\overline{\cal O}^{\rm opt}_{\rm Re,Im}$ according to Eq.(\ref{eq:OOdef}), by substituting $\mathcal{I}$ with $\overline{\mathcal{I}}$ 
and the same for its components $\mathcal{I}_{0,\rm Re, Im}$.

Using these averages, we can extract Re$ F_2$ and Im$F_2$ in an approach similar to Eq.(\ref{eq:reweight}) by setting:
\begin{equation}
\langle \overline{\cal O}^{\rm opt}_{\rm Re,Im}\rangle_{F_2\neq0} = \langle w(F_2) \overline{\cal O}^{\rm opt}_{\rm Re,Im}\rangle_{F_2=0}/\langle w(F_2)\rangle_{F_2=0}
,\end{equation}
where the calculation of the weights $w(F_2)$ has to be performed according to Eq.(\ref{eq:weights}), using the 
not integrated and not averaged components $\mathcal{I}_{0,\rm Re, Im}$. These weights can in fact be calculated, since the data 
set $\mathcal{S}_0$ is generated by Monte Carlo, and the kinematically not reconstructible information of events is available.
Moreover, detector effects can be taken into account in a similar way by calculating $\overline{\cal O}^{\rm opt}_{\rm Re,Im}$ from a data set $\mathcal{S}_0$ with detector simulation included, while using the known generator information in the 
calculation of the weights $w(F_2)$.
\section{Spin-density matrix for $\tau$-lepton pair production in QED to one-loop order}
\label{sec:NLO}
So far we have treated the extraction of the form factor $F_2$ by splitting 
the intensity distribution ${\cal I}$ into a part proportional to $F_2$, and a standard model (one-photon exchange)  part. However, the spin-density matrix for  the pair production process   ${\rm e}^-(\vec k )+ {\rm e}^+(-\vec k )\to \tau^-(\vec p )+ \tau^+(-\vec p )$ 
in the center of mass frame is composed of several contributions up to order  $\alpha^3$ in QED:
\begin{eqnarray}\label{eq:nine}
\chi^{\rm prod}&=& {\alpha^2\sqrt{s-4} \over 16 m_\tau^2 s^{5/2} } \, \Big\{{\cal M}_{\rm tree} + {\cal M}_{{\rm e}-\rm vert} +{\cal M}_{\rm vacpol} \nonumber \\ &&+ {\cal M}_{\tau-\rm vert}  + {\cal M}_{\rm box} + {\cal M}_{\rm  tree}\, \delta_{\rm soft}\Big\}     
\,, \end{eqnarray}
where the prefactor is adjusted to the differential cross section $d\sigma/d\Omega_{\rm cm}$. For notational simplicity, we  omit from now on the four helicity indices $\lambda_+\lambda_-,\lambda_+^\prime\lambda_-^\prime$ on the spin-density matrix $\chi^{\rm prod}$.

In the following we will present explicit  expressions for the six contributions $ {\cal M}_{\rm tree},\dots,
\delta_{\rm soft}$ appearing in Eq.(\ref{eq:nine})  as functions of the
dimensionless Mandelstam variables $s$, $t$ and $u$, which satisfy the constraint $s+t+u=2$. With this convenient choice of variables\footnote{Note that only
in section 3 and appendix A the Mandelstam variable $s$ is dimensionless.}  the initial and final momenta are given in units of the $\tau$-lepton mass $m_\tau = 1776.86\,$MeV by $|\vec k\,| = \sqrt{s}\,m_\tau/2$ and $|\vec p\,| = \sqrt{s-4}\,m_\tau/2$. The invariant squared momentum-transfers,  $t$ and $u$, are expressed in terms of $s$ and the cosine of the scattering angle, $\cos\theta = \hat p \!\cdot\! \hat k$, as:

\begin{eqnarray} t &=& {1\over 2}\Big(2-s+ \sqrt{s^2-4s} \cos\theta\Big) \,,\qquad 2-s <t,u <0\,, \ \nonumber \\ u &=& {1\over 2}\Big(2
  -s-\sqrt{s^2-4s} \cos\theta\Big) \,.\end{eqnarray}

Using these definitions and arranging the prefactor as in Eq.(\ref{eq:nine}), one obtains
from the tree-level one-photon exchange diagram:
\begin{eqnarray}\label{eq:eleven} {\cal M}_{\rm tree}&=& \Big[s+4+(s-4) \cos^2\!\theta\Big]{\bf 1} +(s-4) \sin^2\!\theta \, \vec \sigma_1 \! \cdot \! \vec \sigma_2^*\nonumber \\ &&-2 \Big[ s-4 +(\sqrt{s}-2)^2  \cos^2\!\theta\Big]  \vec \sigma_1 \!\cdot \! \hat p \, \vec \sigma_2^*  \!\cdot \! \hat p\nonumber \\ &&+2(s-2 \sqrt{s})  \cos\theta\Big[  \vec \sigma_1 \! \cdot \! \hat p \, \vec \sigma_2^*  \!\cdot  \!\hat k+ \vec \sigma_1  \!\cdot \! \hat k \, \vec \sigma_2^*  \!\cdot \! \hat p \Big]\nonumber \\ &&-2s \, \vec \sigma_1 \!\cdot \! \hat k \, \vec \sigma_2^*\!\cdot \! \hat k  \,,  \end{eqnarray}
where spin-space 1 belongs to the $\tau^-$ ($\vec\sigma_1 = \vec\sigma_{\lambda_-\lambda_-^\prime}$) and spin-space 2 belongs to the $\tau^+$ ($\vec\sigma_2 = \vec\sigma_{\lambda_+\lambda_+^\prime}$), with the same quantization axis for helicity eigenstates.
The unit-operator ${\bf 1}$ and the products of hermitian spin-operators $O_1 O_2$ get sandwiched  with two-component spin wave functions in the way: $(\chi_1^\dagger O_1\chi_{1'} \!)( \chi_{2'}^t O_2\chi_2^*) \!=\!(\chi_1^\dagger O_1\chi_{1'} \!)( \chi_{2}^\dagger O_2^t\chi_{2'}\!)\\ =(\chi_1^\dagger O_1\chi_{1'} )( \chi_{2}^\dagger O_2^*\chi_{2'})$. The first arrangement of $\chi$-states results from multiplying the original expression (obtained from reducing Dirac-spinors and Dirac-matrices in the diagrammatic amplitude times its complex-conjugated) from the right with the spin-exchange operator $({\bf 1} +\vec \sigma_1 \! \cdot \! \vec \sigma_2)/2$,  and the second or third explain the occurrence of the transposed or complex-conjugated Pauli spin-matrix $\vec \sigma_2^t=\vec \sigma_2^* $. Note that $\vec  \sigma_2^*$ and $- \vec \sigma_2$ are unitarily equivalent due to the relation: $(i\sigma_{2y})\vec  \sigma_2^*(i\sigma_{2y})^{-1} = -\vec \sigma_2$. In this context it is worth to stress that our result for ${\cal M}_{\rm tree}$ in Eq.(\ref{eq:eleven}) agrees with Ref.~\cite{nachtmann} after the substitution $\vec  \sigma_2^*\to -\vec  \sigma_2$.

The first next-to-leading order correction of order $\alpha$ comes from the electronic vertex correction by a photon-loop. Its contribution to the spin-density matrix reads: 
\begin{eqnarray} {\cal M}_{e-\rm vert} &=& 2 {\cal M}_{\rm tree}\, {\rm Re}F_1^{\rm e}(s) \nonumber \\ &=&{\cal M}_{\rm tree}{\alpha \over \pi}\bigg\{ \big(2\xi_{\rm IR}+\ln r\big) \bigg( 1- \ln{s
\over r}\bigg) \nonumber \\ &&- {1\over 2}\ln^2{s \over r} +{3\over 2} \ln{s \over r}+{2 \pi^2 \over 3} -2 \bigg\}\,, 
\end{eqnarray}
where only those terms that do not vanish in the limit $r\to 0$ have been kept, with the very small 
squared mass ratio $r= (m_e/m_\tau)^2=8.271 \cdot 10^{-8}$. The infrared  divergence due to the vanishing photon mass is treated in dimensional regularization and handled by the pole-term:
\begin{equation}\xi_{\rm IR} ={1\over d-4}+{1\over 2} \big(\gamma_E - \ln 4\pi\big) +\ln{m_\tau \over \mu}\,,  \end{equation}
with $d$ the number of space-time dimensions and $\mu$ an arbitrary mass scale. Note that the contribution of the Pauli form factor $F_2^e(s)$ of the electron vanishes in the limit $r\to 0$.

The next radiative correction arises from vacuum polarization in the one-photon exchange:
\begin{equation} {\cal M}_{\rm vacpol} = 2 {\cal M}_{\rm tree} \, \Delta \alpha(\sqrt{s}\,m_t) \,,\end{equation}
with $\Delta \alpha(E_{\rm cm})= {\rm Re}\Pi(E_{\rm cm})$ taken from empirical determinations that include leptonic and hadronic contributions \cite{vacPol}. Ignoring sharp bottomonium resonances, we take an average value of about $\Delta \alpha(10\,{\rm GeV}) \simeq 3.5\%$. We note as an aside that the unique vacuum polarization due to charged leptons $({\rm e}^\mp, \mu^\mp, \tau^\mp)$ reads:
\begin{eqnarray} {\rm Re}\Pi(\sqrt{s}\,m_\tau)^{\rm lept} &=& {2\alpha\over 3\pi} \bigg\{ {s+2 \over s^{3/2}} \sqrt{s-4} \ln {\sqrt{s}+ \sqrt{s-4}\over 2} \nonumber \\ && -{5\over 2} -{2\over s} +\ln s +\ln {m_\tau^2 \over m_e m_\mu}\bigg\}\,,\end{eqnarray}
and this part  amounts to $2.1\%$ at $10\,$GeV. Here, the contributions from electrons and muons are well approximated by leading logarithms plus constants.

The next radiative correction that one has to take into account is the vertex correction at the $\tau$-lepton photon interaction. The corresponding contribution to the spin-density matrix reads:
\begin{eqnarray} {\cal M}_{\tau-\rm vert}&=& 2 {\cal M}_{\rm tree} \, {\rm Re}\big[F_1^\tau(s)+F_2^\tau(s)\big]\nonumber \\ && + (s-4) \Big\{ {\rm Im}F_2^\tau(s) \sqrt{s} \cos\theta \,(\vec \sigma_1\!-\!\vec\sigma_2^*)\!\cdot \!(\hat p \!\times \!\hat k) \nonumber \\ &&  +  {\rm Re}F_2^\tau(s) \Big[ 2 \sin^2\!\theta \, ({\bf 1} -\vec \sigma_1 \! \cdot \! \vec \sigma_2^*)\nonumber \\ && +2\Big(2+(\sqrt{s}-2) \cos^2\!\theta\Big) \vec \sigma_1 \!\cdot \! \hat p \, \vec \sigma_2^*\!\cdot \! \hat p \nonumber \\ && -\sqrt{s} \cos\theta\, \Big( \vec \sigma_1 \! \cdot \! \hat p \, \vec \sigma_2^*  \!\cdot  \!\hat k+ \vec \sigma_1  \!\cdot \! \hat k \, \vec \sigma_2 ^*\! \cdot \! \hat p \Big)\Big] \Big\}\,,\end{eqnarray}
and we recognize here the first correction that is not purely proportional to  ${\cal M}_{\rm tree}$. The photon-loop form factors $F_1^\tau(s)$ and $F_2^\tau(s)$ of the $\tau$-lepton in the time-like region $s>4$ are given by the expressions:
\begin{eqnarray}&& {\rm Re}\big[F_1^\tau(s)\!+\!F_2^\tau(s)\big]= {\alpha\over \pi}
\bigg\{\! \xi_{\rm IR} \!\bigg(\!1 \!+\!{4-2s \over \sqrt{s^2\!-\!4s}} \ln {\sqrt{s}\!+\! \sqrt{s\!-\!4}\over
2} \bigg) \nonumber \\ &&+{3\sqrt{s\!-\!4}\over 2\sqrt{s}} \ln {\sqrt{s}\!+\! \sqrt{s\!-\!4}\over 2} -1+{s - 2\over \sqrt{s^2 \!- \!4 s}}\bigg[ \ln^2{\sqrt{s} \!+\! \sqrt{s \!-\! 4}\over 2}\nonumber \\ && 
 -  \ln{\sqrt{s} \!+ \!\sqrt{s \!-\! 4}\over 2} \ln(s \!-\! 4) +  {\pi^2\over 3}  +  {\rm Li}_2\bigg({s \!-\! 2 \!-\! \sqrt{s^2 \!-\! 4 s}\over 2}\bigg)\!\bigg]\!\bigg\}, \nonumber \\ &&\end{eqnarray}
 and
\begin{equation}F_2^\tau(s) ={ \alpha\over \pi}{1\over  \sqrt{s^2-4s}} \bigg( i \pi -2 \ln {\sqrt{s}+ \sqrt{s-4}\over 2} \bigg) \,,\end{equation}
with Li$_2$ the conventional dilogarithmic function (see also appendix A). Note that $F_2^\tau(0)=\alpha/2\pi = 1.1614 \cdot 10^{-3} $ is the famous Schwinger correction to the lepton anomalous magnetic moment. At 10\,GeV one has a further  suppression of $F_2^\tau(s)$ by a factor $-0.23 +i\, 0.21$.

The last radiative corrections from virtual photon-loops that one has to take into account are the two-photon exchange box diagrams. For the evaluation of the direct and crossed box diagram we follow the decomposition into basic spinorial structures as in section~2 of Ref.~\cite{boxPaper} and introduce eight complex-valued amplitudes:
\begin{eqnarray}&& X_1 = {1\over 2} f_0(s,t) \,-\, [t\to u]\,,  \qquad
                   Y_1 =  {1\over 2} f_0(s,t) + [t\to u]\,, \nonumber \\ &&  X_2 =  f_1(s,t)+  f_2(s,t) \,+\, [t\to u]\,, \nonumber \\ &&  Y_2 =  f_1(s,t)+  f_2(s,t) \,-\, [t\to u]\,, \nonumber \\ && X_3 =  -f_1(s,t)-2  f_3(s,t) \,-\, [t\to u]\,, \nonumber  \\ && Y_3 =  -f_1(s,t) \,-\, [t\to u]\,,  \qquad X_4 = 0\,, \nonumber \\ &&   Y_4 =  -  f_2(s,t) \,-\, [t\to u]\,,   \end{eqnarray}
where the basis functions $f_{0,1,2,3}(s,t)$ are composed of the loop integrals $D_0,\, \overline D_0,\,C_s, \, C_t,\, \overline C_t, \, C_M ,\, B_s-B_M$ and  $B_t-B_M$ (see appendix A for explicit expressions) in the same way as written in Eqs.(7-10) of Ref.~\cite{boxPaper} (setting $M=1$).

Using the complex-valued amplitudes $X_{1,2,3}$ and $Y_{1,2,3,4}$, their individual contributions to the spin-density matrix from the interference of the box 
diagrams with the  one-photon exchange  read:
\begin{equation}  {\cal M}_{\rm box}[X_1]= {\cal M}_{\rm tree} \,{\alpha s \over 2\pi} {\rm Re} X_1\,,    \end{equation}
\begin{eqnarray}  {\cal M}_{\rm box}[Y_1]&=&{\alpha s \over \pi}  \sqrt{s-4}\,\Big\{ -{\rm Im}Y_1 \,(\vec \sigma_1\!-\!\vec\sigma_2^*)\!\cdot \!(\hat p \!\times \!\hat k) \nonumber \\ &&   +  {\rm Re}Y_1 \Big[\sqrt{s} \cos\theta\, {\bf 1} + (2\!-\!\sqrt{s}) \cos\theta \, \vec \sigma_1 \!\cdot \! \hat p \, \vec \sigma_2^*  \!\cdot \! \hat p \nonumber \\ && - \vec \sigma_1 \! \cdot \! \hat p \, \vec \sigma_2^* \!\cdot  \!\hat k- \vec \sigma_1  \!\cdot \! \hat k \, \vec \sigma_2^* \! \cdot \! \hat p \Big] \Big\}\,, \end{eqnarray}

\begin{eqnarray}  {\cal M}_{\rm box}[X_2]&=&{\alpha s \over 4\pi}  \sqrt{s^2-4s}\, \Big\{ -{\rm Im}X_2 \sqrt{s}\,(\vec \sigma_1\!-\!\vec\sigma_2^*)\!\cdot \!(\hat p \!\times \!\hat k)  \nonumber \\ &&  +  {\rm Re}X_2 \Big[\Big({s \over 2}-2\Big)\cos\theta  \sin^2\!\theta ( {\bf 1} -\vec\sigma_1 \! \cdot \!\vec \sigma_2^* )\nonumber \\ && +(\sqrt{s}-2)\Big(2+(\sqrt{s}-2)\cos^2\!\theta \Big)\cos\theta\,\nonumber \\ &&  \times \vec \sigma_1 \!\cdot \! \hat p \, \vec \sigma_2^*  \!\cdot \! \hat p  +s \cos\theta\, \vec \sigma_1 \!\cdot \! \hat k \, \vec \sigma_2^*  \!\cdot \! \hat k \nonumber \\ && -\sqrt{s}\Big(1+(\sqrt{s}-2)\cos^2\!\theta \Big)\nonumber \\ && \times (\vec \sigma_1 \! \cdot \! \hat p \, \vec \sigma_2^* \!\cdot  \!\hat k+ \vec \sigma_1  \!\cdot \! \hat k \, \vec \sigma_2^* \! \cdot \! \hat p)\Big] \Big\} \,, \end{eqnarray}

\begin{eqnarray}  {\cal M}_{\rm box}[Y_2]&=&{\alpha s^2 \over 4\pi}  \sqrt{s-4}\, \bigg\{ -{\rm Im}Y_2 \,(\vec \sigma_1\!+\!\vec\sigma_2^*)\!\cdot \!(\hat p \!\times \!\hat k) \nonumber \\ &&   +  {\rm Re}Y_2 \bigg[ \vec \sigma_1 \! \cdot \! \hat k \, \vec \sigma_2^*  \!\cdot  \!\hat p- \vec \sigma_1  \!\cdot \! \hat p \, \vec \sigma_2^* \! \cdot \! \hat k \nonumber \\ && + \sqrt{s-4} \bigg( {\sin^2\!\theta\over 2}( {\bf 1} +\vec\sigma_1 \! \cdot \!\vec \sigma_2^* )\nonumber \\ &&- \vec \sigma_1 \!\cdot \! \hat p \, \vec \sigma_2^*  \!\cdot \! \hat p   - \vec \sigma_1 \!\cdot \! \hat k \, \vec \sigma_2^*  \!\cdot \! \hat k\nonumber \\ && +\cos\theta (\vec \sigma_1 \! \cdot \! \hat p \, \vec \sigma_2^* \!\cdot  \!\hat k+ \vec \sigma_1  \!\cdot \! \hat k \, \vec \sigma_2^* \! \cdot \! \hat p)\bigg)\bigg]\bigg\}\,, \end{eqnarray} 

\begin{eqnarray}  {\cal M}_{\rm box}[X_3]&=&{\alpha s \over 4\pi}(s-4) \Big\{ -{\rm Im}X_3 \sqrt{s}\cos\theta\,(\vec \sigma_1\!-\!\vec\sigma_2^*)\!\cdot \!(\hat p \!\times \!\hat k)   \nonumber \\ && +  {\rm Re}X_3 \Big[2\sin^2\!\theta \,(\vec\sigma_1 \! \cdot \!\vec \sigma_2^* - {\bf 1} )\nonumber \\ && -2\Big(2+(\sqrt{s}-2)\cos^2\!\theta \Big) \vec \sigma_1 \!\cdot \! \hat p \, \vec \sigma_2^*  \!\cdot \! \hat p  \nonumber \\ && +\sqrt{s} \cos\theta\,  \Big(\vec \sigma_1 \! \cdot \! \hat p \, \vec \sigma_2^* \!\cdot  \!\hat k+ \vec \sigma_1  \!\cdot \! \hat k \, \vec \sigma_2^* \! \cdot \! \hat p\Big)\Big] \Big\} \,, \end{eqnarray}

\begin{eqnarray}  {\cal M}_{\rm box}[Y_3]&=&{\alpha s^2 \over 4\pi}\sqrt{s-4}\, \Big\{ -{\rm Im}Y_3 \,(\vec \sigma_1\!+\!\vec\sigma_2^*)\!\cdot \!(\hat p \!\times \!\hat k)  \nonumber \\ && +  {\rm Re}Y_3 \Big(\vec \sigma_1 \! \cdot \! \hat k \, \vec \sigma_2^* \!\cdot  \!\hat p- \vec \sigma_1  \!\cdot \! \hat p \, \vec \sigma_2 ^*\! \cdot \! \hat k\Big) \Big\}\,,\end{eqnarray}

\begin{eqnarray}  {\cal M}_{\rm box}[Y_4]&=&{\alpha s^2 \over 4\pi}\bigg\{ {\rm Im}Y_4\bigg[{4-s\over \sqrt{s}} \cos\theta\,(\vec \sigma_1\!-\!\vec\sigma_2^*)\!\cdot \!(\hat p \!\times \!\hat k)\nonumber \\ && -\sqrt{s-4}\,(\vec \sigma_1\!+\!\vec\sigma_2^*)\!\cdot \!(\hat p \!\times \!\hat k)\bigg]+ {\rm Re}Y_4 \nonumber \\ &&\times \bigg[ 4\vec \sigma_1 \!\cdot \! \hat k \, \vec \sigma_2 ^* \!\cdot \! \hat k - {\bf 4}+\sqrt{s-4}\,\Big(\vec \sigma_1 \!\cdot \! \hat k \, \vec \sigma_2^*  \!\cdot \! \hat p  \nonumber \\ && -\vec \sigma_1 \!\cdot \! \hat p \, \vec \sigma_2^*  \!\cdot \! \hat k\Big) +(\sqrt{s}-2)^2{\cos\theta\over \sqrt{s}} \Big(\vec \sigma_1 \!\cdot \! \hat p \, \vec \sigma_2^* \!\cdot \! \hat k \nonumber \\ &&  +\vec \sigma_1 \!\cdot \! \hat k \, \vec \sigma_2^*  \!\cdot \! \hat p - 2\cos\theta\, \vec \sigma_1 \!\cdot \! \hat p \, \vec \sigma_2^* \!\cdot \! \hat p\Big) \bigg] \bigg\} \,, \end{eqnarray}
where we have used the relation: $\vec \sigma_1 \!\cdot \!(\hat p \times \!\hat k) \, \vec \sigma_2^* \!\cdot \!( \hat p\times \!\hat k)\\ = \vec \sigma_1 \!\cdot\!\vec \sigma_2^* \sin^2\!\theta - \vec \sigma_1 \!\cdot \! \hat p \, \vec \sigma_2^*  \!\cdot \! \hat p- \vec \sigma_1 \!\cdot \! \hat k \,   \vec \sigma_2^*  \!\cdot \! \hat k+(  \vec\sigma_1 \!\cdot \! \hat p \, \vec \sigma_2^* \!\cdot \! \hat  k \\ + \vec\sigma_1 \!\cdot \! \hat k \, \vec \sigma_2^*  \!\cdot \! \hat p) \cos\theta$.
The total contribution ${\cal M}_{\rm box}$ from the box diagrams is given by the
sum of the seven individual parts corresponding to $X_1, X_2, X_3, Y_1, Y_2, Y_3$ and $Y_4$. According to their construction, the  infrared divergence $ \xi_{\rm IR}$ is contained in the real parts of the
amplitudes $X_1, Y_1, X_2, Y_2$ and $Y_4$. Keeping track of all $ \xi_{\rm IR}$-terms, the total  infrared divergence from the box diagrams is given by the piece:
\begin{equation}  {\cal M}_{\rm box}^{\rm IR}= {\cal M}_{\rm tree} {4\alpha \over \pi}\xi_{\rm IR}\Big[\ln(1-u)-\ln(1-t)\Big] \,, \end{equation}
where half of it comes from Re$X_1$ and the rest arranges itself to the same term proportional to  ${\cal M}_{\rm tree}$.

Finally, there appears at the end of Eq.(\ref{eq:nine}) the correction factor $\delta_{\rm soft}$ from soft photon radiation, which consists of a sum of two pieces: $\delta_{\rm soft}=\delta_{\rm soft}^{\rm (uni)} +\delta_{\rm soft}^{\rm (cm)}$. The universal part:
\begin{eqnarray}\label{eq:soft}\delta_{\rm soft}^{(\rm uni)} &=& {4 \alpha \over \pi} \bigg( \!\ln {m_\tau\over  2\lambda} - \xi_{\rm IR}\!\bigg) \bigg\{ 1+{2-s \over \sqrt{s^2-4s}} \ln {\sqrt{s}+ \sqrt{s-4}\over 2}\nonumber \\ &&-{1\over 2}\ln{s \over r} -\ln(1-t)+ \ln(1-u)\bigg\}   \,,\end{eqnarray}
cancels the infrared  divergence $\xi_{\rm IR}$ from virtual photon-loops and its remainder  depends logarithmically on an infrared cutoff $\lambda$ 
for undetected soft photon radiation. For the numerical studies to be presented in section~\ref{sec:studies} we will choose the value of $\lambda=100\,\mathrm{MeV}$. We estimate in Appendix~C the effect of $\lambda$ on the determination of $F_2$ by scaling it a factor 2 down and up.
The other part is specific for assuming in the center-of-mass frame a small momentum sphere $|\vec \ell\, |<\lambda$
for undetected soft photons, and it reads:
\begin{eqnarray}\label{eq:soft}\delta_{\rm soft}^{(\rm cm)}&=& { \alpha \over \pi} \bigg\{{2\sqrt{s} \over \sqrt{s-4}} \ln {\sqrt{s}+ \sqrt{s-4}\over 2} + \ln{s \over r}+{2 s - 4\over \sqrt{s^2 - 4 s}}\nonumber \\ &&\times \bigg[ \ln^2{\sqrt{s} + \sqrt{s - 4}\over 2}
 -  \ln{\sqrt{s} + \sqrt{s - 4}\over 2} \ln(s - 4)\nonumber \\ &&- {\pi^2\over 6} + {\rm Li}_2\bigg({s - 2 - \sqrt{s^2 - 4 s}\over 2}\bigg)\bigg]  -{1\over 2}\ln^2{s \over r}-{\pi^2 \over 3} \nonumber \\ && + {1\over 2}\ln{1-t \over 1-u} \, \ln{s-1+t u\over s^2}\nonumber \\ &&  + \int_0^1\!{dx \over x} \bigg[{\sqrt{s}(t-1) \over h_t \sqrt{s-4x h_t}} \ln {\sqrt{s} +\sqrt{s-4x h_t}\over  \sqrt{s} -\sqrt{s-4x h_t}} \nonumber \\ &&+\ln{1\!-\!u \over 1\!-\!t}+{\sqrt{s}(1-u) \over h_u \sqrt{s-4x h_u}} \ln {\sqrt{s} +\sqrt{s-4x h_u}\over  \sqrt{s} -\sqrt{s-4x h_u}}\bigg]\!\bigg\},\nonumber \\ && \end{eqnarray}
with the abbreviations $h_t = 1+t(x-1)$ and $h_u = 1+u(x-1)$. Note that the singular behavior with respect to the very small parameter $r=(m_e/m_\tau)^2$ has been extracted and terms vanishing in the limit $r\to 0$ have been dropped. Our calculation of $\delta_{\rm soft}^{(\rm cm)}$ has been guided by the expression in Eq.(39) of Ref.~\cite{radcor} for muon-pair production, performing the limit $r\to 0 $ and interchanging $t\leftrightarrow u$. The last integral $\int_0^1\!dx\, x^{-1}[\dots]$, together with the $(t,u)$-dependent logarithmic term before, is solved by the following  analytical expression:\begin{eqnarray}\label{eq:softIntegral}&&H(s,t,u)= 2 \ln^2(1-t) +\ln\big( s+t-2+t^{-1}\big) \ln{s \over (1-t)^2}  \nonumber \\ &&-  {\rm Li}_2\bigg({(1-t)^2 \over -s t}\bigg)  +2  {\rm Li}_2(t)-4\ln{\sqrt{s}+\sqrt{s-4} \over 2} \nonumber \\ &&\times \ln { \sqrt{-st -(1-t)^2}\,\big(2+\sqrt{-st}+\sqrt{(4-s)t}\,\big) \over \big(1-t+\sqrt{-st}\,\big)\big(1+t+\sqrt{(4-s)t}\,\big)-st -(1-t)^2}\nonumber \\ &&- {\rm Li}_2\Bigg(\bigg[{2t -\sqrt{-st}+  \sqrt{(4-s)t}\over 2+\sqrt{-st}+  \sqrt{(4-s)t}}\bigg]^2\Bigg) \nonumber \\ && + {\rm Li}_2\Bigg(\bigg[{2 -\sqrt{-st}+  \sqrt{(4-s)t}\over 2t+\sqrt{-st}+  \sqrt{(4-s)t}}\bigg]^2\Bigg)\,-\,  [ t\to u].\nonumber \\ &&  \end{eqnarray}
We note that the master integral $I_y^{(2)}$ in appendix A of Ref.~\cite{jan} has been
instrumental in deriving this analytical result for $H(s,t,u)$, and we have performed  also extensive numerical checks. 
At this point one should also mention that the QED radiative corrections to $\tau^- \tau^+$ pair-production have also been 
studied in Ref.~\cite{jadachWas}. However, the entirely different representation of the spin-density matrix and the absence 
of analytical expressions for the virtual photon-loop and soft bremsstrahlung corrections hampers a direct and detailed comparison 
with our results.

In addition to pure QED corrections, we consider also the interference term between one-photon exchange and $Z^0$-exchange. Its contribution to the spin-density matrix in the normalization of Eq.(\ref{eq:nine}) reads:  
\begin{eqnarray}  {\cal M}_{\gamma Z^{0}}&=& {s\, m_\tau^2\over (M_Z\sin\!2\vartheta_W\!)^2} \bigg\{\! \sqrt{s-4}\Big[(\sqrt{s}-2) \cos\theta \, \vec \sigma_1 \!\cdot \! \hat p \, \vec \sigma_2^*  \!\cdot \! \hat p \nonumber \\ &&- \sqrt{s} \cos\theta\, {\bf 1}+ \vec \sigma_1 \! \cdot \! \hat p \, \vec \sigma_2^*  \!\cdot  \!\hat k+ \vec \sigma_1  \!\cdot \! \hat k \, \vec \sigma_2^* \! \cdot \! \hat p \Big]\nonumber \\ && +(1 \!-\! 4\sin^2\!\vartheta_W)\bigg[\! \bigg( \!\sqrt{s\!-\!4}\Big(\!{\sqrt{s}\over 2}(1\!+\! \cos^2\!\theta)-\cos^2\!\theta\Big) \nonumber \\ &&+(s-2\sqrt{s}) \cos\theta \bigg) (\vec \sigma_1\!-\!\vec\sigma_2^*)\!\cdot \!\hat p +\Big(\sqrt{s-4}\cos\theta \nonumber \\ &&+ 2 \sqrt{s}\Big)  (\vec \sigma_1\!-\!\vec\sigma_2^*)\!\cdot \!\hat k \bigg]-{1\over 2}(4\sin^2\!\vartheta_W\!-1\!)^2 {\cal M}_{\rm tree} \bigg\}\,, \nonumber \\ && \end{eqnarray}
where $M_Z=91.19\,$GeV is the $Z^0$ boson mass and $\vartheta_W$ the Weinberg angle with the value $\sin^2\!\vartheta_W = 0.231$ \cite{PDG}. Note that the ratio between the vector coupling and axial-vector coupling of the $Z^0$ to leptons is $1-4\sin^2\!\vartheta_W$, and thus rather small.
\section{Monte-Carlo studies for the measurement of $F_2$}
\label{sec:studies}
In order to estimate the effect of the next-to-leading order QED corrections given in section~\ref{sec:NLO}, we generate data sets 
containing $10^6$ events each for all possible sixteen combinations of the four decay channels:
\begin{equation}
\label{eq:channels}
\pi^-\nu_\tau,\quad\rho^-\nu_\tau,\quad\mathrm{e}^-\nu_\tau\bar\nu_\mathrm{e},\quad\quad\mu^-\nu_\tau\bar\nu_\mu\,,
\end{equation}
and their charge-conjugated counterparts. Since B factories currently offer the best 
way to perform the proposed analysis with real data, we use a center-of-mass energy of 
$\sqrt{s} = m_{\Upsilon(4S)}=10.58\, {\rm GeV}$ for  our numerical studies.
We generate the data sets with a simple sampling and rejecting algorithm and analyze them by using the methods laid out in section~\ref{sec:OO}. Since the kinematics 
of $\tau^-\tau^+$ pair decays cannot be measured completely due to the escaping neutrinos, we have to account 
for this missing information. In the case of $\tau^-$ and $\tau^+$ decaying both semi-leptonically, i.e. 
via $\pi\nu$ or $\rho\nu$, the kinematical information can be reconstructed up to a two-fold ambiguity.
For the measurement of $F_2$, we therefore have to average the intensities over these two 
ambiguous kinematical solutions, as discussed in subsection~\ref{sec:OObar}.

In the case of at least one $\tau$ decaying leptonically ($\tau^-\to\ell^-\nu_\tau\bar\nu_\ell$), the invariant 
mass of the invisible $\nu_\tau\bar\nu_\ell$ system, as well as the directions of the individual neutrinos 
are also unknown. To account for this unknown information, we sample the unknown angles and invariant masses\footnote{With 
a known lepton energy in the center-of-mass frame, the invisible mass square $m^2_{\nu\bar\nu}$ is equally distributed in phase-space.} 100 times for both decays and use the 
mean value of the intensity given in section~\ref{sec:OO} over all kinematically valid samples 
as $\overline{\mathcal{I}}$.
Since every sampled value of the invariant mass results in two solutions for the 
direction of the $\tau$ momenta in the purely semileptonic case, we end up with more than 100 kinematical samples for 
every event on average. This procedure is similar to the one used 
by the Belle collaboration in the measurement of the electric dipole moment $d_\tau$.

Using these techniques, we list in Table~\ref{tab:baseRes} the resolutions obtained for Re$F_2$ and Im$F_2$ in the sixteen different 
combinations of the four decay channels considered in Eq.(\ref{eq:channels}).
The obtained values for $F_2$ are consistent with the input value for every channel.
The resolutions using the true kinematical information yields similar values of
$(\delta{\rm Re}F_2)_\mathrm{true} = 7.0\cdot10^{-4}$ and 
$(\delta{\rm Im}F_2)_\mathrm{true} = 7.2\cdot 10^{-4}$ for all 16 combinations of decay channels considered and we 
observe no significant correlation between the two quantities for the case of fully known final-state kinematics. 

Using $\overline{\cal O}^{\rm opt}_{\rm Re}$, the loss of kinematical information in purely semileptonic channels leads to a decrease 
of the sensitivity of 
only  10\%, while every leptonic decay worsens the sensitivity by more than a factor of two. The loss of kinematical
information affects the resolution of Im$F_2$ for semi-leptonic channels more than for Re$F_2$, while the resolution 
of Im$F_2$ for leptonic channels is less affected by the loss of kinematical information, than in the case of Re$F_2$. 
However, semileptonic channels, especially the $\rho\nu$ channels, still offer a better resolution overall. Since we do not 
observe correlations between Re$F_2$ and Im$F_2$ when using the full kinematical information, the strong correlation 
between Re$F_2$ and Im$F_2$ for purely leptonic channels is introduced by the sampling procedure of kinematics. 
The resolutions in Table~\ref{tab:baseRes} are purely statistical and given for a sample of $10^6$ events in each case. By varying 
the sample size $N$, we have verified that the resolutions scale with $N^{-1/2}$, as it should be.

\begin{table}[h!]
  \begin{center}
    \caption{Results obtained from $10^6$ simulated events for the resolution of Re$F_2$ and Im$F_2$ in the 16 possible combinations of decay channels, and the 
    correlation coefficient $C_{\rm Re,Im}$ between real and imaginary part of $F_2$.}
\medskip    
    \label{tab:baseRes}
    \begin{tabular}{l|l|c|c|c}
    $\tau^-$ mode & $\tau^+$ mode & $10^3\!\cdot\! \delta{\rm Re}F_2$  & $10^3\!\cdot\! \delta{\rm Im}F_2$ & $C_{\rm Re,Im}$\\\hline
    $\pi^-\nu_\tau$  & $\pi^+\bar\nu_\tau$                                              & $0.76$ & $1.15$ & $\phantom{-}0.00886$\\
    $\pi^-\nu_\tau$  & $\rho^+\bar\nu_\tau$                                             & $0.78$ & $0.86$ & $\phantom{-}0.00217$\\ 
    $\pi^-\nu_\tau$  & $\mathrm{e}^+\bar\nu_\tau\nu_\mathrm{e}$                         & $1.45$ & $1.26$ & $\phantom{-}0.00572$\\ 
    $\pi^-\nu_\tau$  & $\mu^+\bar\nu_\tau\nu_\mu$                                       & $1.44$ & $1.24$ & $\phantom{-}0.00200$\\ 
    $\rho^-\nu_\tau$ & $\pi^+\bar\nu_\tau$                                              & $0.78$ & $0.86$ & $-0.00524$\\
    $\rho^-\nu_\tau$ & $\rho^+\bar\nu_\tau$                                             & $0.78$ & $0.81$ & $\phantom{-}0.00509$\\
    $\rho^-\nu_\tau$ & $\mathrm{e}^+\bar\nu_\tau\nu_\mathrm{e}$                         & $1.42$ & $1.29$ & $-0.00396$\\
    $\rho^-\nu_\tau$ & $\mu^+\bar\nu_\tau\nu_\mu$                                       & $1.43$ & $1.29$ & $\phantom{-}0.00596$\\
    $\mathrm{e}^-\nu_\tau\bar\nu_\mathrm{e}$ & $\pi^+\bar\nu_\tau$                      & $1.46$ & $1.30$ & $\phantom{-}0.00428$\\
    $\mathrm{e}^-\nu_\tau\bar\nu_\mathrm{e}$ & $\rho^+\bar\nu_\tau$                     & $1.42$ & $1.26$ & $\phantom{-}0.00084$\\
    $\mathrm{e}^-\nu_\tau\bar\nu_\mathrm{e}$ & $\mathrm{e}^+\bar\nu_\tau\nu_\mathrm{e}$ & $2.42$ & $1.64$ & $\phantom{-}0.63132$\\
    $\mathrm{e}^-\nu_\tau\bar\nu_\mathrm{e}$ &$\mu^+\bar\nu_\tau\nu_\mu$                & $2.61$ & $1.64$ & $-0.69551$\\
    $\mu^-\nu_\tau\bar\nu_\mu$ & $\pi^+\bar\nu_\tau$                                    & $1.45$ & $1.29$ & $\phantom{-}0.00754$\\
    $\mu^-\nu_\tau\bar\nu_\mu$ & $\rho^+\bar\nu_\tau$                                   & $1.42$ & $1.24$ & $\phantom{-}0.00310$\\
    $\mu^-\nu_\tau\bar\nu_\mu$ & $\mathrm{e}^+\bar\nu_\tau\nu_\mathrm{e}$               & $4.08$ & $1.64$ & $-0.88305$\\
    $\mu^-\nu_\tau\bar\nu_\mu$ &$\mu^+\bar\nu_\tau\nu_\mu$                              & $2.18$ & $1.67$ & $\phantom{-}0.51208$
    \end{tabular}
  \end{center}
\end{table}

\subsection{Importance of order $\alpha^3$ corrections}
\label{sec:NLObias}

Since in contrast to the electric dipole moment $d_\tau$, the contributions of the Pauli form factor $F_2$ to the production spin-density matrix are 
CP-even, higher order corrections from QED can mimic the effects of $F_2$ and thus
distort a measurement if not properly taken into account. To estimate the effect 
of higher order QED corrections on the determination of $F_2$, we reanalyze the same 
Monte-Carlo samples as in section~\ref{sec:studies}, but only take the spin-density matrix in tree-level approximation. Thus, the Monte Carlo data set $\mathcal{S}_0$ defined in section~\ref{sec:OO} is also generated using only the tree-level spin-density matrix ${\cal M}_{\rm tree}$. Doing so, 
we find that neglecting higher order 
effects in the analysis leads to a wrong estimation of Re$F_2$ with a value 
of $(-3.46\pm 0.29)\!\cdot\! 10^{-3}$, while the estimation of Im$F_2$ shows a smaller bias of $(0.94\pm 0.26)\!\cdot\! 10^{-3}$.
When using the true kinematical information for every event, this results in
Re$F_2 = (-3.17\pm 0.14)\cdot 10^{-3}$ and Im$F_2= (0.41\pm0.19)\cdot 10^{-3}$. Since the prediction for Im$F_2$  only 
differs  from zero for all channels combined, while for each individual channel it remains consistent with zero for a data-set with $10^6$ events, 
we generate a larger sample of $10^8$ events for the $(\pi^-\nu_\tau)\times(\pi^+\bar\nu_\tau)$ 
mode and perform the analysis, which leads to the result:
\begin{equation}\label{eq:bias}
 {\rm Re}F_2 = (-3.37\pm 0.08)\cdot10^{-3}, \quad {\rm Im}F_2 =  (0.20\pm 0.08)\cdot10^{-3}\,
.\end{equation} 
This bias on the one hand indicates, that corrections of 
order of $\alpha^3$ have to be taken into account for a determination 
of $F_2$. However, since these corrections are of the same order of magnitude as the 
standard model expectation of $F_2^\tau(0)=1.1772 \cdot 10^{-3}$ \cite{eidelman,newHadronic}, 
corrections of order $\alpha^3$ seem to suffice to measure a non-zero value
of $F_2$ for the first time.
With increasing precision, of course, higher order corrections will become 
necessary. 

\section{Summary and conclusion}
\label{sec:conclusions}

We have introduced the method of optimal observables and used it to determine
the Pauli form factor $F_2$ of the $\tau$-lepton. In this study we have
explicitly included  the QED corrections of order $\alpha^3$, as well as
$\gamma Z^0$ interference, and determined the resolutions for Re$F_2$ and Im$F_2$. We have found, that the method gives a similar resolution for $F_2$ in all 
16 combinations of decay channels, when using the true (complete) kinematical information.

By comparing our results for the resolutions of Re$F_2$ and Im$F_2$ with those given in Table~1 of Ref.~\cite{bernabeu}, we find that our 
obtained resolutions are a somewhat larger. However, since the authors do 
not give the total event yield, we cannot perform a more detailed comparison 
of results. In addition, the authors of Ref.~\cite{bernabeu} state, that their resolutions 
for the decay channels $(\pi^-\nu_\tau)\times(\rho^+\bar\nu_\tau)$, $(\rho^-\nu_\tau)\times(\pi^+\bar\nu_\tau)$, and
$(\rho^-\nu_\tau)\times(\rho^+\bar\nu_\tau)$, are worse by factors of 
2 and 4 with respect to the $(\pi^-\nu_\tau)\times(\pi^+\bar\nu_\tau)$ channel and 
thus only ``poorly contribute to increase the precision for the measurement of Re$F_2$''. In contrast to this, we observe a similar resolution for all decay channels, when using 
the true kinematical information, and moreover the $(\rho^-\nu_\tau)\times(\rho^+\bar\nu_\tau)$  channel yields the best resolution at a given luminosity, due
to its
high branching fraction \cite{PDG}.

Since the method proposed in Ref.~\cite{bernabeu} relies on specific 
integrations of the intensity distributions, the detector acceptance 
has to be parametrized as a function of the kinematical variables. 
Otherwise, asymmetric detector 
effects could bias the measurement. Our method, in contrast, only relies 
on an event-by-event detector simulation for the data set $\mathcal{S}_0$ 
to obtain the optimal observables $\overline{\cal O}^{\rm opt}_{\rm Re,Im}$, as defined in section~\ref{sec:OObar}. Such an event-by-event simulation is much
more readily 
available than a parametrization of the detector acceptance. Thus, we believe our
approach to be more robust with respect to detector effects.

Comparing our results to a similar study performed in Ref.~\cite{chen} we find,
that the quoted improvement due to their reconstruction of the neutrino momenta
of a factor of four for $d_\tau$ is mostly due to the different formalism they 
used, rather than due to the resolution of the two-fold kinematical ambiguity. 
As we show in Appendix~B, the ambiguity only 
increases the resolution by a factor 1.61 and 1.06 for Re$(d_\tau)$ and Im$(d_\tau)$.
However, it is difficult to directly compare the resolutions we obtained with the 
ones in Ref.~\cite{chen}, since we did not perform any detector simulation studies.
Additionally, we want to state, that we performed studies separately for the real 
and imaginary parts of the dipole momenta, while the authors of Ref.~\cite{chen}
solely focussed on the real parts. We showed, however, that there are 
differences in the resolutions of real and imaginary part and both parts are 
affected differently by the loss of kinematical information, as can be seen in Table~\ref{tab:baseRes}.

Additionally, we have studied the effect of missing kinematical information 
due to escaping neutrinos, which we have summarized in Table~\ref{tab:baseRes}.
We find that the resolution of $F_2$ is worse for leptonic decay 
channels, since more neutrinos escape in this case. 

Finally, we have studied the effect of QED corrections  of order $\alpha^3$ to the spin-density matrix $\chi^{\rm prod}$ presented in great detail in section~\ref{sec:NLO}. By reanalyzing our generated data sets assuming a tree-level production spin-density matrix ${\cal M}_{\rm tree}$, we find the bias given in Eq.(\ref{eq:bias}) for the measurement of $F_2$. The size of this bias indicates that the inclusion of  $\alpha^3$-corrections allows one to measure the non-zero standard-model value determined in Refs.~\cite{eidelman,newHadronic}, given a correct treatment of photon radiation.
\section*{Appendix A: Loop functions}
\label{sec:loop}
After multiplying with the common factor $16\pi^2$ the complex valued scalar loop-integrals encountered in the evaluation of the two-photon exchange box diagrams (with a light electron and/or heavy $\tau$-propagator) \cite{boxPaper} read in our
notation with dimensionless variables $s$ and $t$:
\begin{equation}D_0 = {2\over s(t-1)}\big( 2\xi_{\rm IR} + \ln s -i \pi\big) \ln{1-t
\over \sqrt{r}}\,,   \end{equation}
\begin{equation}C_s = {1\over s}\bigg( {1\over 2}\ln^2{s\over r}+{\pi^2\over 6}
  - i\pi \ln{s\over r} \bigg) \,,
\end{equation} 
\begin{equation}C_t = {1\over t-1}\bigg[ {\rm Li}_2(t)+ \ln^2{1-t\over \sqrt{r}}
+ \big(2\xi_{\rm IR}+\ln r\big)  \ln{1-t\over \sqrt{r}} \bigg] \,,
\end{equation}
\begin{equation}\overline D_0 = D_0(t\to u)\,, \qquad \overline C_t = C_t(t\to u)\,,\end{equation}
\begin{eqnarray}C_M &=& {2\over \sqrt{s^2-4s}}\bigg[ {\pi^2\over 12} +\ln^2{\sqrt{s}+\sqrt{s-4}\over 2}\nonumber \\ && +{\rm Li}_2\bigg({2-s+\sqrt{s^2-4s}\over 2}\bigg) -i \pi \ln{\sqrt{s}+\sqrt{s-4}\over 2 }\bigg] \,,\nonumber \\ && 
\end{eqnarray}
\begin{equation}B_s-B_M = i \pi -\ln s\,, \quad B_t-B_M={1-t \over t}\ln(1-t) \,, \end{equation}
where ${\rm Li}_2(t)= t \int_1^\infty dx[x(x-t)]^{-1} \ln x$ denotes the dilogarithmic function at argument $t<1$. Moreover, $\xi_{\rm IR} =1/(d-4) +(\gamma_E-\ln 4\pi)/2 +\ln(m_\tau/\mu)$ abbreviates the infrared divergence in dimensional regularization, and $r$ stands for the squared mass ratio $r = (m_e/m_\tau)^2=8.271\!\cdot\!10^{-8} $. We remind that the nomenclature $B$, $C$, and $D$ refers to loop-integrals over two, three, and four propagators, respectively.  
\section*{Appendix B: Results concerning the electric dipole moment}
Since the measurement of the electric dipole moment $d_\tau$ of the
$\tau$-lepton (as performed in Ref.~\cite{belle}) using the optimal observable
method is very  similar to the measurement of the anomalous magnetic dipole moment $a_\tau$ (in units of $e/2m_\tau$), we give also the results of such a study analogous to section~\ref{sec:studies}. The inclusion of an electric dipole moment through a CP-violating form factor $F_3(s)$ adds the following structure:
\begin{equation}
\frac{F_3(s)}{2m_\tau} \sigma^{\mu\nu}\gamma^5 q_\nu
\end{equation}
to the Dirac-matrix $\Gamma^\mu$  in Eq.(\ref{eq:QEDgamma}). The
conversion formula of $F_3(s)$ to the electric dipole moment  $d_\tau$, commonly
given in units of $e\mathrm{cm}$,  reads:
\begin{equation}
d_\tau = \frac{e}{2 m_\tau}F_3(0)  = 5.44\cdot 10^{-15}\,e\mathrm{cm}
\cdot F_3(0)\,.\end{equation}
Clearly, in $\tau^-\tau^+$ pair-production one can only obtain 
$F_3(s)$ for $s > 4 m_\tau^2$.

The resulting resolutions for the extraction of $F_3$ are 
given in Table~\ref{tab:baseResEDM}. Without loss of kinematical 
information, the resolutions are similar and take the values 
of $(\delta{\rm Re} F_3)_\mathrm{true} = 0.93 \cdot 10^{-3}$ and $(\delta{\rm Im}F_3)_\mathrm{true} = 0.52 \!\cdot\! 10^{-3}$. We observe no significant correlations between Re$F_3$ and Im$F_3$. A reanalysis of the generated data neglecting next-to-leading order QED effects, similar to 
section~\ref{sec:NLObias}, yields no bias for the extraction of $F_3$. 
This is expected, since any contribution from $F_3$ is CP-odd, while 
QED contributions are CP-even both at tree-level and one-loop order.

\begin{table}[h!]
  \begin{center}
    \caption{Results from $10^6$ simulated events for the resolutions of Re$F_3$ and Im$F_3$ in the 16 possible combinations of decay channels.}

    \medskip
    \label{tab:baseResEDM}
    \begin{tabular}{l|l|c|c}
    $\tau^-$ mode & $\tau^+$ mode & $10^3\!\cdot\!\delta{\rm Re}F_3$  & $10^3\!\cdot\!\delta{\rm Im}F_3$\\\hline
    $\pi^-\nu_\tau$  & $\pi^+\bar\nu_\tau$                                              & $1.50$ & $0.55$ \\
    $\pi^-\nu_\tau$  & $\rho^+\bar\nu_\tau$                                             & $1.11$ & $0.57$ \\ 
    $\pi^-\nu_\tau$  & $\mathrm{e}^+\bar\nu_\tau\nu_\mathrm{e}$                         & $3.57$ & $1.00$ \\ 
    $\pi^-\nu_\tau$  & $\mu^+\bar\nu_\tau\nu_\mu$                                       & $3.47$ & $1.00$ \\ 
    $\rho^-\nu_\tau$ & $\pi^+\bar\nu_\tau$                                              & $1.11$ & $0.57$ \\
    $\rho^-\nu_\tau$ & $\rho^+\bar\nu_\tau$                                             & $1.07$ & $0.57$ \\
    $\rho^-\nu_\tau$ & $\mathrm{e}^+\bar\nu_\tau\nu_\mathrm{e}$                         & $2.96$ & $1.00$ \\
    $\rho^-\nu_\tau$ & $\mu^+\bar\nu_\tau\nu_\mu$                                       & $2.95$ & $1.00$ \\
    $\mathrm{e}^-\nu_\tau\bar\nu_\mathrm{e}$ & $\pi^+\bar\nu_\tau$                      & $3.69$ & $1.06$ \\
    $\mathrm{e}^-\nu_\tau\bar\nu_\mathrm{e}$ & $\rho^+\bar\nu_\tau$                     & $3.09$ & $0.95$ \\
    $\mathrm{e}^-\nu_\tau\bar\nu_\mathrm{e}$ & $\mathrm{e}^+\bar\nu_\tau\nu_\mathrm{e}$ & $20.3$ & $1.73$ \\
    $\mathrm{e}^-\nu_\tau\bar\nu_\mathrm{e}$ &$\mu^+\bar\nu_\tau\nu_\mu$                & $10.9$ & $1.71$ \\
    $\mu^-\nu_\tau\bar\nu_\mu$ & $\pi^+\bar\nu_\tau$                                    & $3.65$ & $1.05$ \\
    $\mu^-\nu_\tau\bar\nu_\mu$ & $\rho^+\bar\nu_\tau$                                   & $3.02$ & $0.95$ \\
    $\mu^-\nu_\tau\bar\nu_\mu$ & $\mathrm{e}^+\bar\nu_\tau\nu_\mathrm{e}$               & $11.4$ & $1.72$ \\
    $\mu^-\nu_\tau\bar\nu_\mu$ &$\mu^+\bar\nu_\tau\nu_\mu$                              & $13.4$ & $1.70$ 
    \end{tabular}
  \end{center}
\end{table}
\section*{Appendix C: Variation of the soft photon cutoff}
In order to study the effect of the soft photon cutoff $\lambda$ in Eq.(\ref{eq:soft}) we 
generate $10^8$ events for the $(\pi^-\nu_\tau)\times(\pi^+\bar\nu_\tau)$ mode employing the values $\lambda= 50$MeV and $\lambda=200$MeV and analyze these data-sets with our method, assuming a value of $\lambda=100$MeV, which was used in our studies above. 
Apart from the different values of $\lambda$, the models used for data generation  and analysis are equal.
Doing so, leads to the following bias in the determination of $F_2$:
\begin{align}
\text{Re}F_2^{\lambda=50\text{MeV}}  &= (2.53\pm0.05)\!\cdot\!10^{-3},\\
\text{Im}F_2^{\lambda=50\text{MeV}}  &= (-0.16\pm0.05)\!\cdot\!10^{-3},\\
\text{Re}F_2^{\lambda=200\text{MeV}} &= (-1.75\pm0.06)\!\cdot\!10^{-3},\\
\text{Im}F_2^{\lambda=200\text{MeV}} &= (0.03\pm0.07)\!\cdot\!10^{-3}
.\end{align}
Since these values are of similar size as the values found in Eq.(\ref{eq:bias}),
a correct treatment of photon radiation turns out to be crucial 
for a measurement of $F_2$.
For $F_3$ we again find no bias in this study, as expected.

Since the value of $\lambda$ plays a role, it has to be chosen properly to represent 
the energy resolution of an actual experiment. The treatment of bremsstrahlung in 
${\rm e}^- {\rm e}^+ \to \tau^- \tau^+$ beyond the soft photon approximation goes beyond the 
scope of the present paper since such a calculation should be tailored to the 
specific experimental conditions for undetectable photon radiation including detector 
efficiencies and various other effects.

\section*{Acknowledgements}
We thank R.~Karl, S.~Paul, J.~Portoles and A.~Rostomyan for useful discussions.\\ This work
has been supported in part by DFG (Project-ID 196253076 - TRR 110) and NSFC.

\end{document}